\documentclass[fleqn,twoside]{article}
\usepackage{espcrc2}

\usepackage{graphicx}
\usepackage[figuresright]{rotating}

\def\gta{ \lower .75ex \hbox{$\sim$} \llap{\raise .27ex \hbox{$>$}} }
\def\lta{ \lower .75ex\hbox{$\sim$} \llap{\raise .27ex \hbox{$<$}} }

\newcommand{\AmS}{{\protect\the\textfont2
  A\kern-.1667em\lower.5ex\hbox{M}\kern-.125emS}}

\title{The XMM/BeppoSAX observation of Mkn 841}

\author{P.O. Petrucci\address[laog]{Laboratoire d'Astrophysique de
 Grenoble,
 BP 43, 38041 Grenoble  Cedex 9, France},
 C. Cabanac\addressmark[laog],
 G. Henri\addressmark[laog],
 L. Maraschi\address[oab]{Osservatorio Astronomico di Brera,
 Via Brera 28, 02121 Milano, Italy},
 P. Ferrando\address[cea]{Service d'Astrophysique, DSM/DAPNIA/SAp, CEA
  Saclay, 91191 Gif-sur-Yvette Cedex, France},
 G. Matt\address[oar]{Dipartimento di Fisica,
  Universit\`a ``Roma tre'',
 via della Vasca Navale 84, I-00046 Roma, Italy},
 M. Mouchet\address[luth]{LUTH, Observatoire de Paris, Section de
  Meudon,
 92195 Meudon Cedex, France},
 C. Perola\addressmark[oar],
 S. Collin\addressmark[luth],
 A.M. Dumont\addressmark[luth],
 F. Haardt\address[ins]{Universit\`a dell'Insubria,
 Via Lucini 3, 22100 Como, Italy},
 L. Koch-Miramond\addressmark[cea]}

\begin{document}

\begin{abstract}
Mkn 841 has been observed simultaneously by XMM and BeppoSAX in January
2001. Due to operational contingency, the 30ks XMM observation was
split into two parts, separated by about 15 hours. We first report the
presence of a narrow iron line which appears to be rapidly variable
between the two pointings, requiring a non-standard interpretation. We
then focus on the analysis of the broad band (0.3-200 keV) continuum
using the XMM/EPIC, RGS and SAX/PDS data.  The Mkn 841 spectrum is well
fitted by a comptonization model in a geometry more photon-fed than a
simple slab geometry above a passive disk. It presents a relatively
large reflection (R$\gta$2) which does not agree with an apparently
weak iron line. It also show the presence of a strong soft excess well
fitted by a comptonized spectrum in a cool plasma, suggesting the
presence of a multi-temperature corona.  \vspace{1pc}
\end{abstract}

\maketitle

\section{The data}
The 2001 XMM observation of Mkn 841 was split into two parts: the first
one, called hereafter OBS1, was done the 13th of February and lasted 12
ks. The second observation, hereafter OBS2, was done the 14th of February
and lasted 15 ks. The two observations were separated by about 15 hours.

The EPIC-pn camera was operated in Small Window mode with the thin
aluminium filters. All the EPIC and RGS event files were reprocessed
using SASv5.4.0. The EPIC spectra were extracted from a 40 arcsec window
with PATTERN $\leq$ 4 and $\leq$ 12 for the PN and MOS respectively.
There was no pile up and the background was low during the whole
observation.

A simultaneous BeppoSAX observation was done between the 11th and 14th of
February with a total net exposure time of 90 ks for the MECS.
 
\begin{figure}[!h]
\includegraphics[width=\columnwidth]{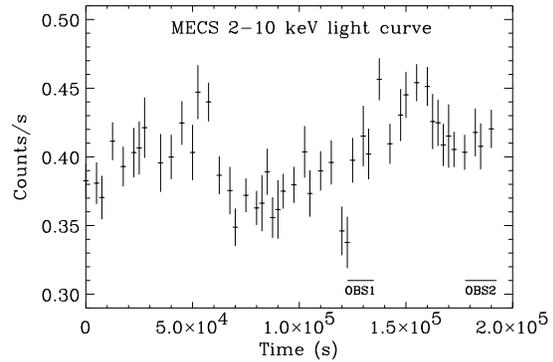}
\caption{MECS light curves (bins of 2500 s). The periods of the two XMM
observations are indicated on the figure.}
\label{fig1}
\end{figure}
The MECS light curve is reported in Fig. 1 with a binning of 2500 s. The
periods of the two simultaneous XMM observations are also reported in
this figure. The integrated XMM-Newton flux of Mkn 841 over the 2-10 keV
range was roughly the same for the two EPIC-pn observations at about 1.4
10$^{-11}$erg.s$^{-1}$.cm$^{-2}$. The (2-5 keV)/(5-10 keV) hardness ratio
of the MECS was consistent with no spectral variability during the
observation. We thus integrated the MECS and PDS spectrum over the entire
observation
\begin{figure*}[t]
\begin{tabular}{cc}
\includegraphics[height=\columnwidth,angle=-90]{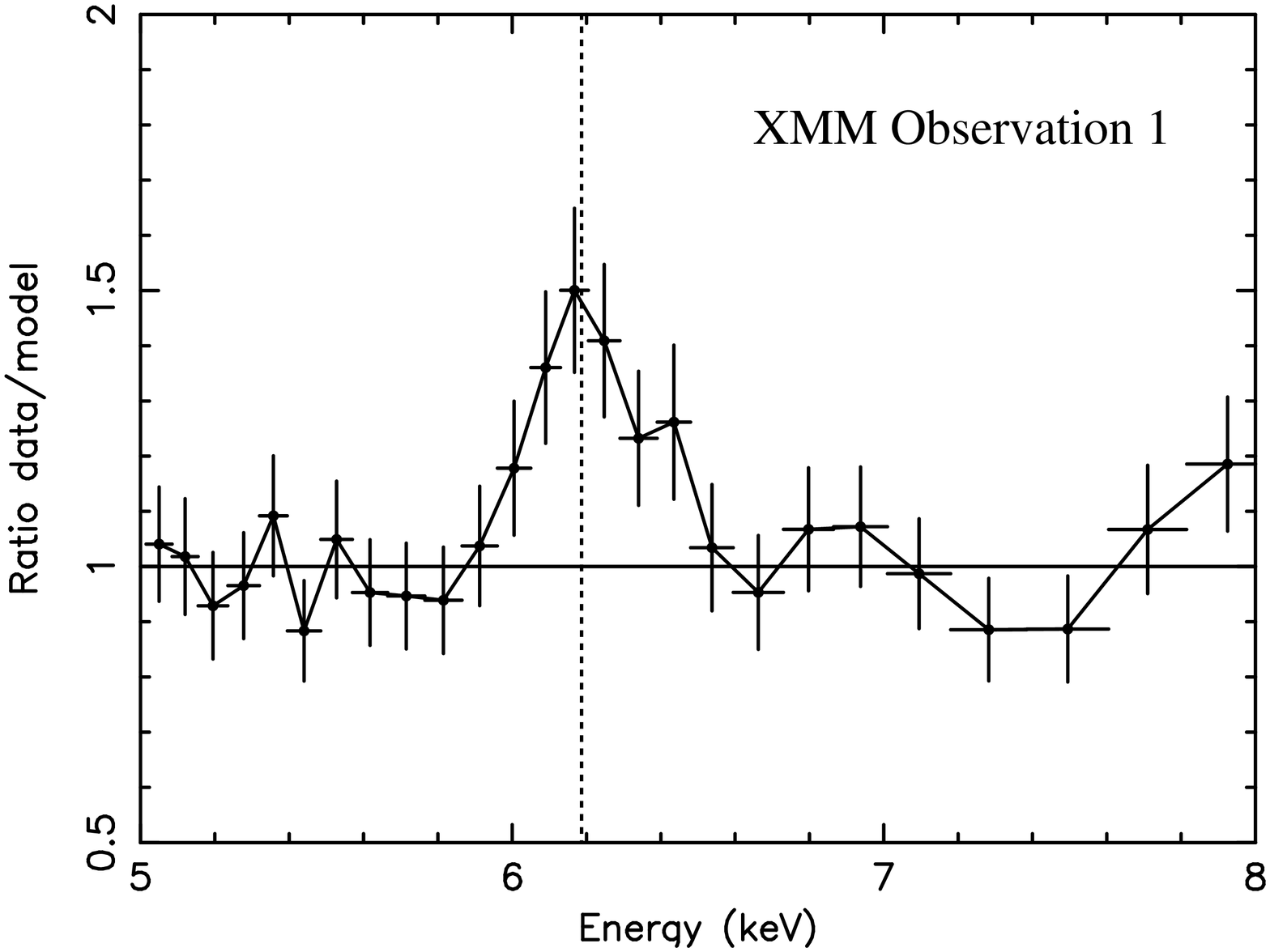}&\includegraphics[height=\columnwidth,angle=-90]{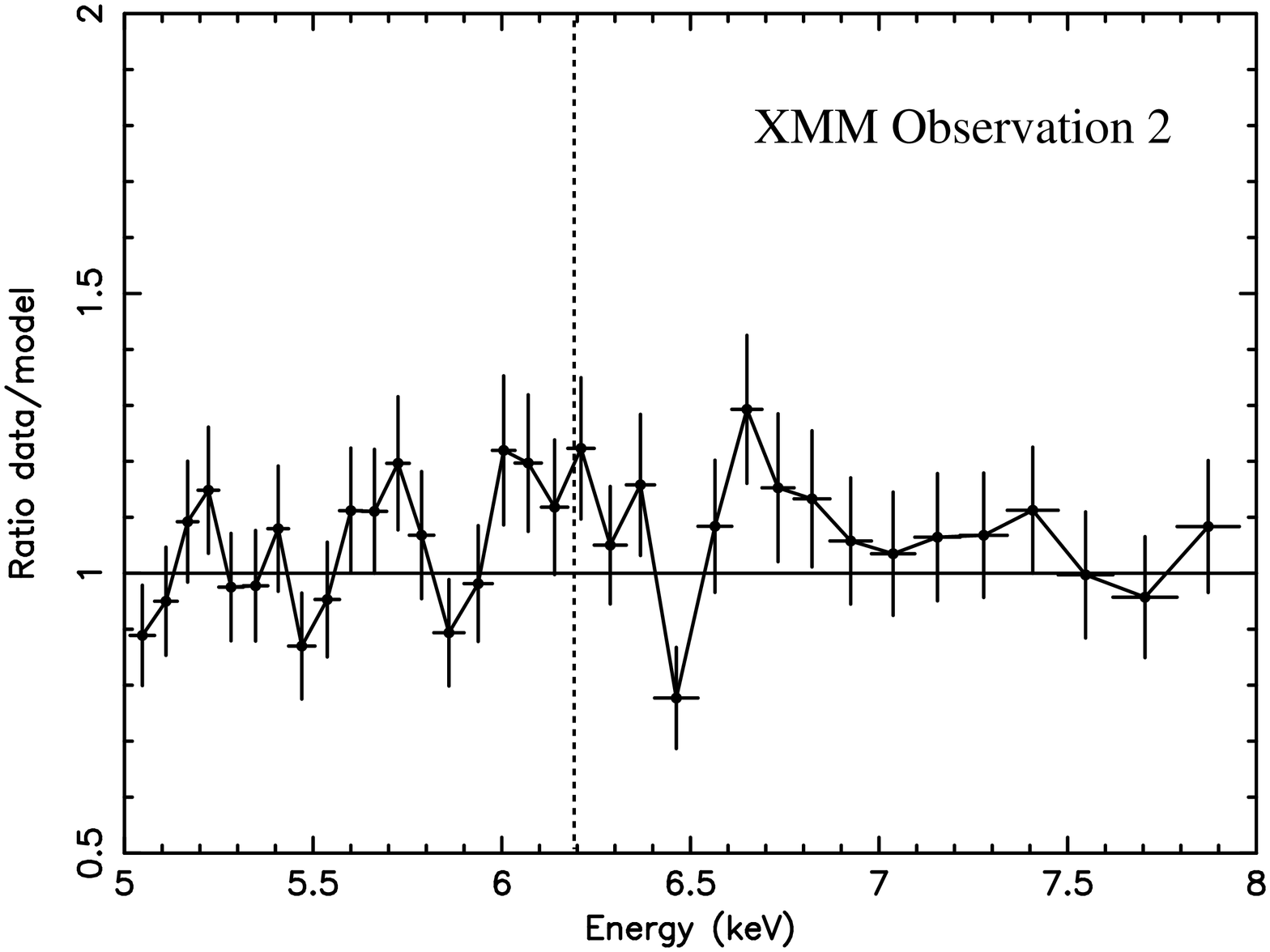}\\
\includegraphics[width=\columnwidth]{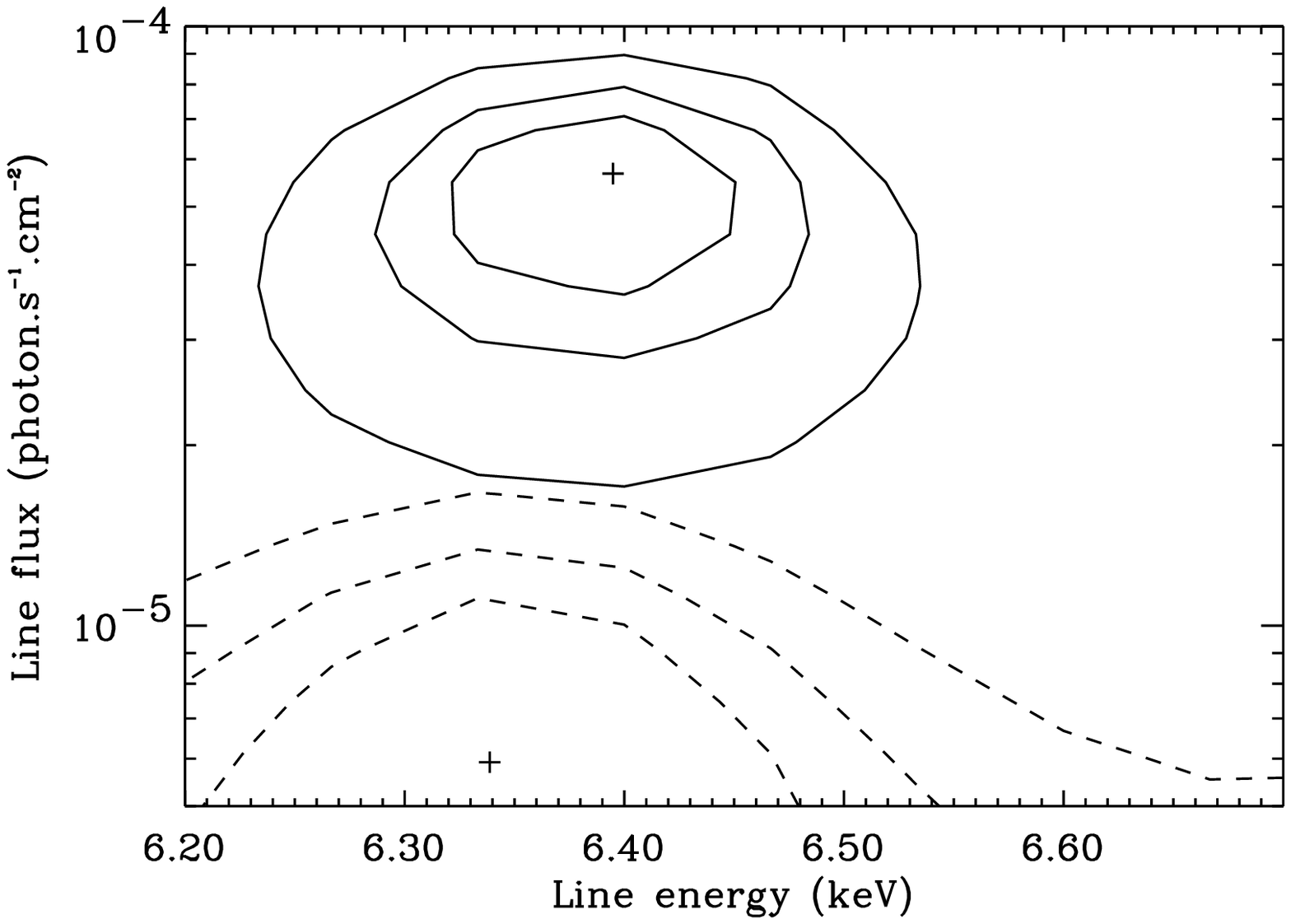}&\begin{minipage}{\columnwidth}
\vspace*{-6cm}
\caption{{\bf Top:} ratios data/model (without fitting the line) for OBS1 and
  OBS2. The {\sc pexrav} fit was done using the EPIC-pn and PDS data
  simultaneously. The vertical dashed line indicates the position of the
  6.4 keV neutral iron line in the source rest frame.  {\bf Bottom:}
  contour plots (68\%, 90\% and 99\% confidence level) of the line flux
  vs. the line energy during the line flux maximum of OBS1 (solid line)
  and the total OBS2 (dashed line).}
\end{minipage}
\end{tabular}
\label{fig2}
\end{figure*}

\section{A rapidly variable narrow iron line}

We detect a rapidly variable narrow neutral Fe K$\alpha$ line (Petrucci
et al. 2002, hereafter P02) between OBS1 and OBS2. The line flux reaches
a maximum during the first observation and is significantly reduced in
the second one (cf. Fig. 2). The continuum shape and flux, instead, keep
roughly constant between the two pointings. The reflexion appears to be
relatively large (cf. Table 1) with R$>$1 while the line EW is relatively
small.

This results cannot be easily explained in the standard cold reflection
model framework where the narrow line component is supposed to be
produced far from the central black hole and is then not expected to be
rapidly variable. Some (not conclusive) explanations (cf. P02 for
details) are:
\begin{itemize}
\item Micro-flares in the inner part of the disc (\cite{nay01},
  \cite{yaq01})
\item Warped concave disc (\cite{bla99})
\end{itemize}

\begin{table*}[!htb]
\caption{Best fits of the continuum above 2 keV. {\sc compha} is the
comptonization code in slab geometry of Haardt (\cite{haa94}). {\sc
  ion. disc} is the 
constant density ionised disc model of Ross \& Fabian (\cite{ros93})}
\begin{center}
\begin{tabular}{ccccccc}
\hline
Model & $\Gamma$ & $E_c/kT_e$ (keV) & $\tau$ & $\xi$ & $R$ & $\chi^2$/dof \\
\hline
\sc{pexrav} & 2.04$_{-0.05}^{0.09}$ & 115$_{-60}^{+270}$ & - & - &
2.5$_{-0.9}^{+1.1}$ & 330/301\\
\sc{compha$^*$} & - & 180$_{-20}^{+15}$ & 7$_{-2}^{+2} \times 10^{-2}$ & -
& 4.7$_{-1.5}^{+1.0}$ & 335/301\\
\sc{Ion. disc} & 1.91$_{-0.03}^{+0.03}$ & - & - & 1.8$_{-0.2}^{+0.1}$ &
0.9$_{-0.4}^{+0.2}$ & 341/299\\ 
\hline
\end{tabular}\\
\end{center}
$^*$ the soft photon temperature is fixed to 10 eV
\end{table*}

\section{The 2-200 keV continuum}
Fitting the XMM/PN data with a simple power law give a good fit to the
continuum, and show very small spectral and flux variations between the
two XMM observations.  

To better constrain the continuum we then fit the XMM/PN and SAX/PDS
simultaneously above 2 keV. We use different types of model: 1) a simple
cut-off power law + reflection ({\sc pexrav} model of {\sc xspec}), 2) a
comptonization model in slab geometry ({\sc compha}, \cite{haa94}), 3)
the ionized disc model of Ross \& Fabian ({\sc ion. disc} \cite{ros93})

Best fit results have been reported in Table 1.
\begin{itemize}
\item The 3 models give acceptable fits but the ionized disc fits give
  the larger reduced $\chi^2$.
\item The addition of a high energy cut-off in the {\sc pexrav} model is
  required at more than 98\% (following the F-test)
\item The comptonization fit agrees with a more photon-fed geometry (the
  Compton parameter is equal to 0.25) than a slab corona above a passive
  disk suggesting e.g. intrinsic disc emission.
\item We found large R values with the first two models ({\sc pexrav} and
  {\sc compha}) while the {\sc Ion Disc} gives a more reasonable value
  R$\sim$1 (however cf. next section).
\end{itemize}

\begin{figure}[!h]
\includegraphics[height=\columnwidth,angle=-90]{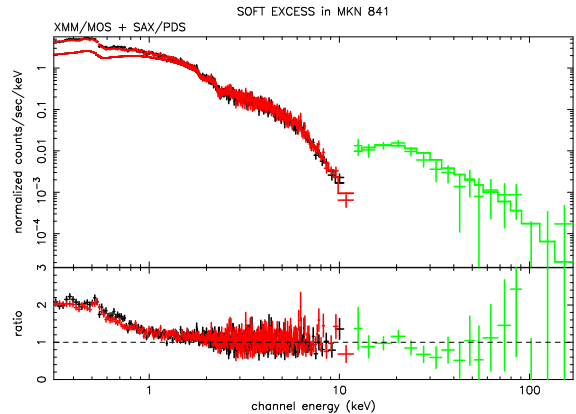}
\caption{Best fit of the XMM/MOS (OBS 1 and OBS 2) and SAX/PDS data with
  {\sc pexrav} when fitting the data above 2 keV only and then
  extrapolating down to 0.3 kev. A soft excess is clearly visible.}
\label{fig3}
\end{figure}

\begin{table*}[t]
\caption{Best fit of the soft excess, fixing the primary continuum
parameters to those reported in Table 1. For {\sc compha} and {\sc
  comptt} the soft 
photon temperature is fixed to 10 eV.}
\begin{center}
\begin{tabular}{cccccccc}
\hline
Continuum & Soft excess & N$_h^*$ & kT$_{bb}$ (eV) & $\Gamma$ &  $kT$ (keV) & $\tau$ & $\chi^2$/dof \\
\hline
\sc{pexrav} & Black body & - & 115$_{-1}^{+2}$ & - & - & - & 1158/623\\
       & Power law & 6.0$_{-0.9}^{+1.4}$ & - & 4.3$_{-0.1}^{+0.1}$ & - & - & 777/623\\
       & {\sc comptt} & 5.4$_{-1.5}^{+1.0}$ & - & - & 1.3$_{-0.6}^{+3.0}$ &
       3.5$_{-1.1}^{+2.0}$ & 771/622\\
\hline
\sc{compha} & Black body & - & 120$_{-2}^{+2}$ & - & - & - & 1243/623\\
       & Power law & 6.5$_{-1.0}^{+1.1}$ & - & 4.2$_{-0.1}^{+0.1}$ & - & - & 781/623\\
       & {\sc comptt} & 5.1$_{-1.5}^{+1.0}$ & - & - & 0.9$_{-0.6}^{+3.0}$ &
       4.6$_{-1.8}^{+1.5}$ & 771/622\\
\hline
\sc{Ion. Disc} & Black body & - & 115$_{-3}^{+2}$ & - & - & - & 1753/623\\
       & Power law & 6.5$_{-1.0}^{+1.0}$ & - & 4.3$_{-0.1}^{+0.1}$ & - & - & 863/623\\
       & {\sc comptt} & 6.6$_{-1.0}^{+0.9}$ & - & - & 60$_{-5}^{+5}$ &
       $<0.44$ & 864/622\\
\hline
\end{tabular}\\
\end{center}
$^*$ in excess to the galactic one
\end{table*}

\section{The soft excess}

Extrapolating the best fit models obtained above 2 keV to low energy, a
soft excess is always clearly present in the data (cf. Fig. 3). We found
that:
\begin{itemize}
\item It does not vary between the two XMM observations
\item It represents $\sim$40-50\% of the 0.3-2 keV flux
\item No strong emission features are present in the RGS data suggesting a
 thermal continuum origin. Absorption features however (also observed in
 the EPIC data) reveal the presence of a warm absorber 
\end{itemize}
We fit the soft excess (fixing the primary continuum parameters to their
best fit values reported in Table 1) with different models. Results are
reported in Table 2. 

A simple black body is a very bad fit of the soft excess. A better fit is
obtained with a comptonization spectrum (model {\sc comptt} of {\sc
  xspec}) suggesting the soft excess to be the hard tail of the disc
emission comptonized in a cool (few keV) corona. We note that the {\sc
  ion. disc} model akways gives significantly worse broad band fits (cf.
Table 2) and is thus ruled out by the data.

\section{Conclusion and perspective}

Mrk841 has been observed simultaneously by BeppoSAX and XMM. The main
results of this study can be summarize as follows (P02, Petrucci et al.
\cite{pet03}):
\begin{itemize}
\item We detect the presence of a rapidly variable but narrow and neutral
  iron line 
\item Our best fits suggest a photon fed geometry and give a very large
 reflection composant   
\item A strong soft excess is present in the data and is better fit by a
 comptonized spectrum 
\end{itemize}
The line variability is surprising and requires a non-standard
interpretation. It has to be confirmed by new observations to better
constrain its origin. A XMM observation of 75 ks has been accepted in
priority B for AO3.

The presence of a large reflection component suggests either an
anisotropy of the primary emission (as already proposed by Bianchi et al.
\cite{bia01}), or a large reflecting material solid angle (e.g. concave
disc, \cite{bla99}).  The last case would agree with a photon-fed
geometry suggested by the thermal comptonization fits. Moreover, in this
framework, the line variability could be explained by the precession of
the warped disc (\cite{har00}). Such geometry appears thus as an
attractive explanation for the X-ray spectral shape and variability
behavior of Mkn 841.

These data also suggest the presence of a multi-temperature corona, the
cooler regions ($\sim$ 2 keV) producing the soft excess while the hotter
ones ($\sim$ 200 keV) produce the hard X-ray continuum.


\begin{thebibliography}{9}
\bibitem{bla99} Blackman 1999, MNRAS, 306, L25
\bibitem{bia01} Bianchi, S., Matt, G., Haardt, F. et al.\ 2001,
A\&A, 376, 77
\bibitem{haa94} Haardt 1994, PhD dissertation, SISSA, Trieste
\bibitem{har00} Hartnoll \& Blackman 2000, MNRAS, 317, 880
\bibitem{nay01} Nayakshin, S.~\& Kazanas, D.\ 2001, ApJL, 553,
L141
\bibitem{pet02} Petrucci et al. 2002, A\&A, 388, L5 (P02)
\bibitem{pet03} Petrucci et al. 2003, in preparation
\bibitem{ros93} Ross \& Fabian 1993, MNRAS, 261, 74
\bibitem{yaq01} Yaqoob, T.\ 2001, Proc. of ``New Century of X-ray
Astronomy'', Japan, eds.H. Inoue, H. Kunieda, ASP Conf. Series
\end{thebibliography}
\end{document}